\documentclass[preprint,authoryear,12pt]{elsarticle}
\usepackage{graphicx}
\usepackage{amsmath,amssymb}
\usepackage{multirow}
\usepackage{array}

\begin{document}

\providecommand{\umin}{\ensuremath{u_\textrm{min}}}
\providecommand{\umean}{\ensuremath{u_\textrm{mean}}}
\providecommand{\uminhat}{\ensuremath{\hat{u}_\textrm{min}}}
\providecommand{\umeanhat}{\ensuremath{\hat{u}_\textrm{mean}}}
	\begin{frontmatter}

	\title{Complexity Measures and Concept Learning}


\author[econ,coco]{Andreas D. Pape}
\author[psych,coco]{Kenneth J. Kurtz}
\author[bioen,coco]{Hiroki Sayama}

\address[econ]{Corresponding Author.  Department of Economics, Binghamton University, PO Box 6000, Binghamton, NY 13905. \texttt{apape@binghamton.edu}. 1-607-777-2660.}
\address[psych]{Department of Psychology, Binghamton University, PO Box 6000, Binghamton, NY 13905. \texttt{kkurtz@binghamton.edu}}
\address[bioen]{Departments of Bioengineering and Systems Science and Industrial Engineering, Binghamton University, PO Box 6000, Binghamton, NY 13905.  \texttt{sayama@binghamton.edu}}
\address[coco]{Collective Dynamics of Complex Systems Research Group, Binghamton University, PO Box 6000, Binghamton, NY 13905}

	\begin{abstract}
		The nature of concept learning is a core question in cognitive science. Theories must account for the relative difficulty of acquiring different concepts by supervised learners.  
	   For a canonical set of six category types, two distinct orderings of classification difficulty have been found. 
	  One ordering, which we call paradigm-specific, occurs when adult human learners classify objects with easily distinguishable characteristics such as size, shape, and shading. The general order occurs in all other known cases: when adult humans classify objects with characteristics that are not readily distinguished (e.g., brightness, saturation, hue); for children and monkeys; and when categorization difficulty is extrapolated from errors in identification learning. 
	The paradigm-specific order was found to be predictable mathematically by measuring the logical complexity of tasks, i.e., how concisely the solution can be represented by logical rules.  
	
	  However, logical complexity explains only the paradigm-specific order but not the general order. 
	  Here we propose a new difficulty measurement, information complexity,
that calculates the amount of uncertainty remaining when a subset of the dimensions are specified. This measurement is based on Shannon entropy. 
	  We show that, when the metric extracts minimal uncertainties, this new measurement predicts the paradigm-specific order for the canonical six category types, and when the metric extracts average uncertainties, this new measurement predicts the general order.
	  Moreover, for learning category types beyond the canonical six, we find that the minimal-uncertainty formulation correctly predicts the paradigm-specific order as well or better than existing metrics (Boolean complexity and GIST) in most cases.
	 
	\end{abstract}


	\end{frontmatter}

\section{Introduction}

In a canonical classification learning experiment, human learners are tested on the six possible categorizations that assign eight examples (all possibilities of three binary-valued dimensions) to two equal-sized classes  (Shepard, Hovland, \& Jenkins, 1961).\nocite{shepard1961learning}
  These classification problems, commonly referred to as the SHJ types, have been instrumental in the development and evaluation of theories and models of category learning. Learning is easiest for Type I in which the classes can be distinguished using a simple rule on a single dimension--e.g. all large items are category A and all small items are category B.  Learning is most difficult for Type $VI$ in which the two classes cannot be distinguished according to any set of rules or statistical regularities.  The remaining types ($II-V$) are intermediate in difficulty.  (Table \ref{tab:SHJmappings} provides a complete description of the six mappings.)

These experiments yield a well-known ordering with a particular pattern across the intermediate types: Type $II$ (a logical XOR rule on two dimensions) is learned faster than Types $III - V$, which are learned at the same speed. An update to this traditional SHJ ordering based on a review of the existing literature and a series of new experiments reveals that Type $II$ does not differ from Types $III - V$ except under particular instructional conditions that encourage rule formation or attention to particular dimensions (Kurtz, Levering, Romero, Stanton, \& Morris, 2012).\nocite{shepard1961learning,nosofsky1994comparing,kurtzRevising}

While this ordering (with or without the recent update) is generally what researchers associate with the SHJ types, there also exists a set of results across a wide variety of learning circumstances in which an entirely different ordering occurs. Specifically, the intermediate types separate into an ordering as follows: $I < IV < III < V < II < VI$. Of particular note is the difficulty in learning Type $II$ (along with the non-equivalence of Types $III-V$). There are four separate cases that yield results consistent with this ordering: first, stimulus generalization theory, which generates a prediction of the ordering of the classification problems based on the frequency of mistakes (pairwise confusions) in learning unique labels (i.e., identification learning) for each item \citep{shepard1961learning}; second, stimuli comprised of integral dimensions \citep{garner1974processing} that are difficult for the learner to perceptually analyze and distinguish, such as brightness, hue, and saturation \citep{nosofsky1996learning}; third, learning by monkeys (Smith, Minda, \& Washburn, 2004); fourth, learning by children (Minda, Desroches, \& Church, 2008).\nocite{minda2008learning}

Since this less well-known ordering occurs across such far-reaching circumstances, we will refer to it as the \emph{general order}; and since the well-known SHJ ordering is only found in one specific learning setting (adult humans learning to classify separable stimuli), we will refer to it as the \emph{paradigm-specific order}. We acknowledge that for some readers, it may seem counterintuitive to dissociate the ordering they are most familiar with from the ordering we designate as general, but in fact it makes good sense to do so.

To provide further detail about the evidence for the general ordering, it has been shown that the results for learning the SHJ types with integral-dimension stimuli  fully match the general order, i.e. $I < IV < III < V < II < VI$ \citep{nosofsky1996learning}.
	Since this also corresponds to stimuli generalization theory, these results
	are interpreted as reinforcing Shepard et al.'s (1961) view that stimuli 
	generalization theory predicts ease of learning unless a process of attention or abstraction can be applied by the learner.  

The settings with non-adult or non-human learners match an important characteristic of the general order, that $II$ is found to be more difficult than Types $III-V$, while there is only some support for the $IV < III < V$ ordering.
In the cross-species research \citep{smith2004category}, four rhesus monkeys were tested on a modified version of the SHJ six types.  
	The core finding is that Type $II$ was more difficult for the monkeys to learn than Types $III-V$ (which the authors elect to average across in their reporting).  
In the developmental work \citep{minda2008learning}, the researchers modified the SHJ task 
	to be age-appropriate for children of ages 3, 5, and 8. 
	Only Types $I-IV$ were tested: Type $II$ was the most difficult to learn (consistent with the general rather than the paradigm-specific order).  
	No significant difference between Types $III$ and $IV$ was observed, however it appears that the researchers did not 
	evaluate the interaction between age of children and their performance on Types $III$ and $IV$.  
	From the mean accuracy data, it can be seen that the children show increasingly good performance on Type $III$ with
	age and increasingly poor performance with age on Type $IV$.  While we do not have access to statistical support, 
	the available evidence is consistent with the younger children learning Type IV more easily than Type $III$ (as in the general ordering).

There are two general classes of explanation in the psychological literature on category learning that have been successfully applied to the SHJ types.
Mechanistic models, which are implemented in computational simulations of trial-by-trial learning, have been used to explain the paradigm-specific order (i.e. Kurtz, 2007; Love, Medin, \& Gureckis, 2004)
\nocite{love2004sustain, kurtz2007divergent}
and some have been shown to account for both the paradigm-specific and general orders \citep{kruschke1992alcove,nosofsky1996learning,PapeKurtzGEB}.
The other approach is based on the use of formal metrics to measure mathematical (logical) complexity
(Feldman, 2000, 2006; Goodman, Tenenbaum, Feldman, \& Griffiths, 2008; Goodwin and Johnson-Laird, 2011; Lafond, Lacouture, \& Mineau, 2007; Vigo, 2006, 2009, 2013).
These models heretofore account only for the paradigm-specific order. \nocite{feldman2000,feldman2006algebra,goodwin2011mental,lafond2007complexity,vigo2006note,Vigo2009,Goodman2008,vigo2013gist}

We put forth a mathematical complexity metric, information complexity, which can account for, on one hand, the paradigm-specific order and, on the other hand, the general order, with a single change in the formula from a \emph{min} to a \emph{mean} operator.  
Our metric calculates the Shannon entropy \citep{shannon2001mathematical} in a classification problem when a subset of the dimensions are specified.  
The \emph{min} operator identifies the subsets of dimensions which provide the most information (and thus leave the \emph{minimal} uncertainty): this applies to the paradigm-specific order, in which sophisticated learners can observe separable dimensions and
may employ abstraction or attention with regard to these dimensions.
On the other hand, the \emph{mean} operator averages over subsets of dimensions, and, correspondingly, it applies to the general order, in which learners are less sophisticated or unable to separate dimensions.  
The logic of this correspondence is described in greater detail in Section \ref{sec:theory} (Theory).
Among complexity accounts of learning behavior, this new measurement has the advantage of being an analytical function exclusively of observable parameters, i.e. it does not require a heuristic to calculate \citep{feldman2000} nor does it require the fitting parameters to data \citep{vigo2013gist}.

In Section \ref{sec:theory}, we describe the background of information theory and define the metric.  
In Section \ref{sec:results}, we evaluate the metric's prediction of learning behavior.  
	In Section \ref{sub:shj_results}, 
		we demonstrate the metric's ability to predict the paradigm-specific and general orders of the SHJ tasks, 
			as well as show it successfully predicts quantitative error rates. 
	In Section \ref{sub:beyond_shj_results}, 
		we demonstrate the metric's ability to predict the paradigm-specific ordering on classification learning tasks beyond SHJ 
			as well or better than the existing metrics (Boolean complexity and GIST) in all cases but one. 
		We also show it successfully predicts the quantitative error rates.  
		The general order setting has not been tested beyond SHJ: this section also, therefore, provides predictions for those future experiments.  

\section{Theory}\label{sec:theory}

In this section, we first provide a comparison of the existing metrics in the literature, which rely on logical complexity, and Shannon entropy, which provides the foundation of our metric.  Then, we formally introduce our metric and explain its components.

\subsection{Logical Complexity versus Information Complexity} 
\label{sub:measuring_complexity}


\emph{Logical complexity} characterizes the length of the shortest description of a system.  In an SHJ-style classification, the `system' in question is a particular categorization.  
Feldman's Boolean complexity \citep{feldman2000}
is a type of logical complexity metric, but there are others, such as Kolmogorov (algorithmic) complexity, which is the length of the shortest program to produce a certain output \citep{li2008introduction}.  These are all related in the sense that they are attempting to construct a minimal set of logical rules that describe a system or process or categorization.  The measurement of Boolean complexity begins with the `disjunctive normal form' of a classification, which is the most verbose way to describe the classification as a set of values connected by AND and OR, i.e. ``(small AND dark AND circle) OR (large AND dark AND circle).''  Then heuristics are applied to eliminate redundant elements, and the metric is defined as the final, minimal number of remaining logical literals.

Vigo's GIST \citep{vigo2013gist} is not strictly a logical-complexity metric, but also incorporates aspects models based on selective attention, such as GCM, ALCOVE, and SUSTAIN \citep{nosofsky1986attention,kruschke1992alcove,love2004sustain}.  GIST stands for ``Generalized Invariance Structure Theory.''  The term ``invariance'' refers to distilled elements of a category when a dimension is suppressed or ignored.  Objects that appear multiple times under these conditions are considered `invariant.'\footnote{An earlier version of this model, CIT \citep{Vigo2009}, involves perturbing the dimensional value and considering objects that remain in the category, which is a more natural notion of `invariance.'}  The essence of the GIST metric is that the more invariants, the easier a category is to learn; the fewer invariants, the harder.  
%
Invariants are somewhat similar to the notion of redundant elements in Boolean complexity.

\emph{Information complexity} characterizes the amount of information or uncertainty in a system.
%
The most commonly used metric, Shannon information entropy \citep{shannon2001mathematical}, measures how much information an observer can gain from one observation of a system: the higher the Shannon entropy, the more unpredictable.  
To explain Shannon entropy, we begin with the formal definition and then consider an example of a fair and unfair coin.

First, the definition of Shannon entropy.
Consider a single random variable $X$, which can take on finitely many values $\{x_1,    x_2, \ldots, x_n\}$, where value $x_i$ occurs with probability $p_i$.  Then the Shannon entropy of $X$ is given by:
\begin{align}
	H(X) &=  \sum_{i=1}^{n} p_i \cdot I(p_i), \nonumber \\ 
\textrm{where }	I(p) &= - \log_2 p  \nonumber
\end{align}
In the information theory literature, the function $I(p)$ is called the ``self-information'' of an event which occurs with probability $p$.  $I(p)$ can be interpreted as the \emph{informativeness} of observing such an event, i.e. how much surprise the event engenders.  Taking that interpretation as given, 
$H(X)$ is then the probabilistically-weighted sum of the informativeness of all possible outcomes of $X$.  This means $H(X)$ is the expected value of informativeness of $X$.  

$H(X)$ can be interpreted as uncertainty.  Why?  $I(p)$ is the \emph{ex-post} `surprise' engendered by a single observation of an event, which means that $H(X)$ is the expected \emph{ex-ante} `surprise.'  That does it mean to \emph{expect surprise}?  Consider the opposite: suppose one expects to not be surprised by an outcome.  When one expects not to be surprised, one can be said to be fairly certain.  Therefore, if one expects to be surprised, one can be said to be uncertain.  Therefore, $H(X)$ can be interpreted as `uncertainty.' 

To further clarify the interpretation of $I(p)=-\log_2 p$ as the informativeness of an event which occurs with probability $p$,
note that over the range of valid probabilities,
 $-\log_2 p$ goes to $\infty$ as $p$ approaches $0$, 
 it is $1$ at $p=\frac{1}{2}$, and it is $0$ at $p = 1$.
%
%
%
This reflects the idea that an event that occurs with vanishingly small probability is very surprising when it occurs, so is very informative.  On the other hand, an event that occurs with certainty (i.e. with probability $1$), is completely uninformative: what happens is what one is sure would happen.

Our metric is founded on this way of measuring uncertainty.
Uncertainty is maximized when all events are equally probable.  Intuitively, this rests on symmetry, and the fact that making one event less likely increases the informativeness of that event, but also necessarily makes the other event more likely, decreasing the informativeness of that other event.  In the extreme, a \emph{certain} event has zero informativeness, because the informativeness of the certain event ($=0$) is multiplied by the total available probability of one, and the other event is ignored.  Mathematically, this arises from maximizing uncertainty with straightforward calculus.  For example, in the case of two events, $p I(p) + (1-p) I(1-p)$ is maximized at $p=.5$.

\medskip

Consider the uncertainty of a fair, then unfair coin.
Suppose $X'$ is the flip of a fair coin, 
where the outcomes are ``heads'' and ``tails''
and $p_{\textrm{heads}}=p_{\textrm{tails}}=.5$.
The informativeness of each ``heads'' or ``tails'' is $-\log_2 \frac{1}{2} = 1$, as mentioned above.  Therefore, the information entropy is 
$H(X') = \left(.5 \cdot 1 + .5 \cdot 1 \right) = 1$.
This can be interpreted as: a flip of a fair coin reveals, on average, exactly one binary `bit' of information.
On the other hand, an unfair coin reveals, on average, less than one bit of information:
Suppose $X''$ is an unfair coin, where ``heads'' and ``tails'' occur with probabilities $.75$ and $.25$ respectively. Then
\begin{align}
H(X'') &= 	.25 \cdot \left(-\log_2 .25 \right) + .75 \cdot \left(-\log_2 .75 \right) \nonumber \\
	&= \left( 	.25 \cdot 2 + .75 \cdot 0.41	\right) \nonumber \\
\implies H(X'') & \approx .81 < 1  \nonumber 
\end{align}  This means that it yields, on average, \emph{less} than one bit of information. 
In fact,
the more unfair the coin, the less the \emph{average} flip reveals: the intense surprise of the unlikely event is smothered by its vanishingly small probability.  This culminates at the extreme of a maximally unfair coin, $X'''$, whose `flips' are certain and therefore are Shannon entropy-free: 
\begin{align}
	H(X''')  &= 	0 \cdot \left(-\log_2 0 \right) + 1 \cdot \left(-\log_2 1 \right) \nonumber 
\end{align}
The fact that $\lim_{p\rightarrow 0} \left( - p \log_2 p \right) = 0$ establishes that the first term is zero, so $H(X''') = 0 + 0 = 0$.

\bigskip


\subsection{The Information Complexity Metric} 
\label{sub:our_metric}

Our complexity metric is based on Shannon information entropy; as mentioned above, it is founded on an aggregation of informativeness.  The metric encodes:
If the object's characteristics for some subset of the dimensions are specified, how much uncertainty is left in the categorization?   In this way, we construct an aggregate measure of the informativeness of each dimension and each set of dimensions.

Formally, let a classification task be formulated as a binary function $f(x) \to \{A, B\}$, where $A$ and $B$ are two categories and $x$ is a multidimensional vector with $d$ dimensions, so $x = (x_1, x_2, \ldots, x_d)$.  
Let $X$ be a finite subset of the real line and is a list of possible values that the entries of $x$ can take on, so $x\in X^d$.  In SHJ, $d_\textrm{SHJ} = 3$, $X_\textrm{SHJ} = \{0,1\}$, and there are six functions $f$ (when symmetric cases are aggregated), one associated with each Problem Type $I-VI$.


The information complexity metric rests on calculating the average remaining uncertainties in classification after entriess of $x$ are specified to certain values.  It is convenient to define these quantities by grouping them by the number $n \in \{0,1,\ldots, d\}$ of dimensions which are specified.  We begin constructing the metric by considering different ways to partition the stimuli into subsets when $n$ dimensions are specified.\footnote{A \emph{partition} of a set $Y$ is a set of non-overlapping subsets of $Y$ that together contain every element of $Y$.  I.e. $\{Y_1, Y_2, \ldots, Y_n\}$ is a partition of $Y$ if: $Y_1 \subseteq Y$ for all $i$, $Y_i \cap Y_j = \emptyset$ for all $i\neq j$, and $\bigcup_i Y_i = Y$.  }  We define $S(n)$ as the set of all such $n$-dimension partitions.
Formally, we define $S(n)$ as:
\begin{eqnarray}
S(n) &=& 
\Bigg\{ \;
\Big\{
\big\{(x_1, x_2, \ldots, x_d) \; | \; (x_{i_1}, \dots, x_{i_n}) = (b_1, \ldots, b_n)  \big\}
\; \Big| \;
\{b_1, \ldots, b_n\} \in X^n \Big\}\nonumber \\
&& \quad \, \Big| \quad \{i_1, \ldots, i_n\} \subseteq \{1, \ldots, d \}
\Bigg\} \nonumber
\end{eqnarray}
  In SHJ, $X_\textrm{SHJ}=\{0,1\}$, and, if we let $\star$ denote an unspecified `digit' of $x$, then, for example,
\begin{align}
	S_\textrm{SHJ}\left(1\right) &=
					\left\{
						\begin{array}{ll}
						\{ \left(0,\star,\star\right) ,&		\left(1,\star,\star\right) \}, \\
						\{ \left(\star,0,\star\right) ,&		\left(\star,1,\star\right) \}, \\
						\{ \left(\star,\star,0\right) ,&		\left(\star,\star,1\right) \}
						\end{array}
						\right\}  \nonumber
\end{align}
Note that, as described earlier, each element of $S_\textrm{SHJ}(1)$ is a partition of $X^d$; for example, the second element represents the set
\[\big\{\;\{(0,0,0), (1,0,0), (0,0,1), (1,0,1)\}, \; \{(0,1,0), (1,1,0), (0,1,1), (1,1,1)\}\;\big\}.\]
As this example demonstrates, this \emph{partitions} the set of all $|X|^d=2^3=8$ stimuli into two subsets: one in which the second element is $0$ and one in which the second element is $1$.  

Now we wish to consider the average remaining uncertainty for each element of $S(n)$.  We do so by defining a vector $U(n)$, which has a non-negative real number for each element of $S(n)$. Each element of $S(n)$ is a partition of the stimulus space, so the entry in $U(n)$ associated with a particular partition represents the average uncertainty associated with that partition, when one considers `learning' which subset in that partition one falls in.
So $U_{SHJ}(1)$ also has three elements, one for each element of $S_{SHJ}\left(1\right)$, and the first entry of $U_{SHJ}(1)$ is the remaining uncertainty after learning, in this case, the first digit of the stimulus.
The values of the entries of $U(n)$ vary with the problem type determined by function $f$: i.e. Type $I$ and Type $VI$ classification problems have different amounts of uncertainties for the same subsets of specified dimensions.
Formally, 
\begin{eqnarray}
	U(n) &=& 
	\left\{ 
\frac{1}{|s|}\sum_{y \in s } Q(y)
		\; \middle| \; s \in S(n)	
	 \right\} , \nonumber
\end{eqnarray}
where $Q(y)$ is defined as
\begin{eqnarray}
Q(y) &=&
	\sum\limits_{a \in \{A,B\}}
			J\big(
				p\left(
					f(x) = a | x \in y
					\right)
				\big)  \nonumber
\end{eqnarray}
with $J(p) = p \cdot I(p) = - p \log_2 p$.
$Q(y)$ is the function that calculates the entropy (remaining uncertainty) in classification within $y$, i.e., when entries of $x$ in the focal $n$ dimensions are set to a particular set of values; for example, when the second dimension in SHJ is set to 1. Then each element of $U(n)$ is a simple average of $Q(y)$ over $y$, e.g., the average uncertainty when the second dimension in SHJ is set to 0 or 1.

Then our metric is defined as:
\begin{align}
	u_G(n) &= G( U(n) ) \label{uGdefn} 
\end{align}
where the function $G$ is an \emph{aggregation function} that produces the amount of relevant overall uncertainty from $U(n)$. There are two forms of $G$ that are of interest here, \emph{min} and \emph{mean}, which we discuss further in Section \ref{sec:results}.  As Equation \ref{uGdefn} suggests, we denote the two metrics as \umin{} and \umean{}.

Given this definition, $u_G(0)$ represents the average remaining uncertainty when no dimensions are determined, so $S(0)$ has one element, which is a set that contains the whole set $X^d$.  $u_G(0) = 1$ if both categories, A and B, are equally present in the stimulus space. 
Clearly, $u_G(0)$ 
must equal or exceed $u_G(n)$ for all $n>0$.
 On the other extreme, $u_G(d)$ gives the smallest value because the least unpredictability remains when all dimensions are determined.  In other words, $S(d)$ is a collection of singleton sets, where each set contains precisely one element of the whole set $X^d$.
In the SHJ series of experiments, observing all dimensions uniquely defines the category, so $u_G(d) = 0$.  (In principle, one could consider categorization learning in which the categories have some unresolvable uncertainty, in which case $u_G(d) > 0$.)

For completeness, the statement of the information complexity metric in a single expression is:
\begin{align}
u_G(n) = G \Big( & \Big\{ 
	\frac{1}{|X^n|} \sum_{\{b_1, \ldots, b_n\} \in X^n} 
		\sum_{a \in \{A,B\}}
				J\big(
					p\left(
						f(x) = a 
						\middle| 
						(x_{i_1}, \dots, x_{i_n}) = (b_1, \ldots, b_n) 
						\right)
					\big) 
			\nonumber \\
& \quad\Big| 
	\{i_1, \ldots, i_n\} \subseteq \{1, \ldots, d\}
 \Big\}  \Big) \nonumber
\end{align}

We aggregate this metric over all possible values of $n$, i.e. all possible numbers of specified dimensions.  This aggregate information complexity metric is:
\begin{equation}
\hat{u}_G = \sum_{n = 0}^d u_G(n) = \sum_{n = 0}^d G( U(n) )  \nonumber
\end{equation}
We consider $\hat{u}_G$ to be a measure of the overall information complexity of a classification task.

\begin{table}
	\begin{center}
	\begin{tabular}{|c|c|}
		\hline
		\textbf{A} & \textbf{B} \\ \hline
		$\left\{0,0,0\right\}$  &  	$\left\{0,1,0\right\}$  \\
		$\left\{0,0,1\right\}$  &  	$\left\{0,1,1\right\}$  \\
		$\left\{1,1,0\right\}$  &  	$\left\{1,0,0\right\}$  \\
		$\left\{1,1,1\right\}$  &  	$\left\{1,0,1\right\}$  \\ \hline
	\end{tabular}
	\end{center}
	\caption{SHJ Type II}
	\label{tab:shjtype2}
\end{table}

\subsection{Comparison to Boolean Complexity and GIST} 
\label{sub:comparison_to_boolean_complexity_and_gist}

To better understand the four metrics considered here---\uminhat{}, \umeanhat{}, Boolean complexity, and GIST---we compare how they evaluate SHJ Type II.  SHJ Type II is depicted in Table \ref{tab:shjtype2}, where the stimuli are depicted as three-dimensional binary vectors.  As can be seen in this table, the `rule' which defines Type II is, if the first two dimensions match, then it is in category A; if they do not, then it is in category B.  This implies that the third dimension has no effect on the categorization.

First, let us consider the Boolean complexity representation of this categorization.  Boolean complexity begins with a maximally verbose logical description of the elements of A, reduces the statement, and measures its length.  The notation used by Feldman (and Boole) is the following: we let $a$, $b$, and $c$ represent the claims ``The first digit is 1,'' ``the second digit is 1,'' and ``the third digit is 1,'' respectively.  Then $ab$ represents the claim ``$a$ AND $b$,'' $a + b$ represents the claim ``$a$ OR $b$,'' and $a'$ represents the claim ``not $a$.''  Given this notation, the most verbose description of the elements of A in SHJ Type II is: $abc + abc' + a'b'c + a'b'c'$.  The most compact representation is $ab + a'b'$, which, translated back into verbal claims, is ``the first and second digits are both one or the first and second digits are both zero,'' which the reader can confirm corresponds to the fundamental rule describing Type II above.  Feldman's Boolean complexity in this case involves simply counting the `literals' (i.e. claims $a$, $b$, or $c$) that appear in the most compact representation.  In the most compact representation, $a$ appears twice and $b$ appears twice for a total of four literals, so the Boolean complexity of SHJ Type II is $4.$

Second, let us consider how GIST evaluates this categorization.  GIST considers `invariants' based on ignoring or `binding' dimensions one by one.  First, GIST constructs what Vigo calls a ``structural manifold,'' in which it calculates the proportion of stimuli in $A$ are invariants when a particular dimension is `bound.'  Consider binding the third dimension.  When the third dimension is bound (ignored) \emph{every} stimulus in category $A$ is an invariant, so the proportion associated with the third dimension is $1$.  By contrast, binding either of the first two dimensions produces \emph{no} invariants, to the proportion associated with those dimensions are zero.   Therefore the structural manifold for SHJ Type II is $(0,0,1)$.  (In general, these proportions can lie between zero and one.)  The manifold $(0,0,1)$ indicates that the first two dimensions are the most useful to observe to categorize the object (i.e. `contain the most information,') while the third is the least useful.  GIST transforms the structural manifold into a single metric by taking the square root of the sum of squared entries, and calling the resulting value $\widehat{\Phi}$, which can be considered the value of the GIST metric.\footnote{This is not precisely correct.  In fact, $\widehat{\Phi}$ is then transformed so that smaller $\widehat{\Phi}$ corresponds to larger values and a functional form is applied which affects the magnitude.  However, since the reversed order of $\widehat{\Phi}$ is preserved, the value of $\widehat{\Phi}$ captures the essence of GIST.}
  In this case, $\widehat{\Phi}\left(SHJ\, II\right) = \sqrt{ 0^2+0^2+1^2 }=1$.

Third and finally, let us consider our information complexity metrics \uminhat{} and \umeanhat{}.  As described above, they are identical but for aggregation at the dimension level; accordingly, we are able to consider both at the same time.  The metric is constructed at each dimension level $d$  from zero to three (in this case), where $d$ denotes the number of dimensions which are fixed.  Consider $d=1$.  If one dimension is known to be (say) zero, what is the probability that a particular stimulus is in (say) category $A$?  The corresponding probabilities are $(.5, .5, .5)$; that is, there are as many category $A$ stimuli with a zero as the first entry as category $B$ stimuli, and so on.  This corresponds to maximal uncertainty in each case.  Therefore the vector of uncertainties at dimension level $d=1$ is $(1,1,1)$.  This vector of uncertainties is called $U(1)$ in the notation above.\footnote{Uncertainty is equal to the $J$ associated with category $A$ plus the $J$ associated with category $B$; so $J(.5) + J(.5)= 2 \cdot (- .5) \cdot \log_2 .5 = 2 \cdot (- .5) \cdot ( - 1) = 1$.}  Now consider $d=2$.  If two dimensions are known, what is the uncertainty about the stimulus category $A$?  First, consider the first two dimensions.  When the first two dimensions are known, then the category is known with certainty.  That implies that the uncertainty associated with the first two dimensions is zero.  Now consider the case when the first and third dimension are known.  In this case, the category is completely unknown (there as many stimuli with first and third dimensions equalling, for example, $(0,0)$, in the category $A$ as $B$) and so uncertainty is maximized at $1$.  This same argument applies when the second and third dimensions are known (uncertainty is maximized).  Therefore, the vector of uncertainties associated with $d=2$ is $U(2) = (0,1,1)$.  Having considered the non-trivial cases of $d=1$ and $d=2$, let us now consider the trivial cases.  Consider $d=3$.  Uncertainty is minimized when all dimensions are observed, so the associated uncertainty `vector' is $U(3) = (0)$.  (Why does this vector have one entry?  Because there is only one way to divide the stimuli into subsets where all dimensions are observed: one subset for each stimulus.)  Finally, consider $d=0$.  Clearly, uncertainty is maximized when no dimensions are observed, so the uncertainty `vector' $U(0) = (1)$.  (This vector has one entry because there is only one way to divide the stimuli into subsets there no dimensions are observed: all stimuli are in one subset.)

Now we apply the aggregation method, which reveals the difference between the version of the metric which reveals the paradigm-specific order and the one which yields the general order.  The \emph{mean} aggregator assumes that at each dimension level $d$, the `best' way to divide up the stimuli cannot be chosen.  So the mean aggregator yields $mean (1,1,1)$ for $d=1$, which is $1$, and $mean(0,1,1)$ for $d=2$, which is $.67$; and obviously also selects $mean (1) = 1$ for $d=0$ and $mean (0) = 0$ for $d=3$.  Therefore $\umeanhat{} = 1+1+.67+0 = 2.67$.  By contrast, the \emph{min} aggregator assumes that the `best' way to divide up the stimuli at each dimension level can be chosen.  So the $min$ aggregator yields $min (1,1,1) = 1$ for $d=1$ (no difference between $min$ and $mean$) and $min(0,1,1) = 0$ for $d=2$, compared to $mean(0,1,1) = .67$ for the $mean$ aggregator.  This obviously corresponds to the idea that the `best' way to break up the stimuli at the $d=2$ level is to divide up the stimulus space by considering the first two dimensions instead of any other pairing, and there is where the value of the min aggregator emerges.  So $\uminhat{} = 1+1+0+0 = 2$.\footnote{These values also appear in Table \ref{tab:paradigm-order-outcome} (for \uminhat{}) and Table \ref{tab:general-order-outcome} (for \umeanhat{}).} 

In an important way, GIST can be thought of as a method which is in two ways the `opposite' of the information complexity metric.
GIST looks for stimuli which become identical when dimensions are ignored.  By contrast, in our information complexity metric, we fix the value of a dimension---which is the `opposite' of ignoring that dimension---and look for remaining \emph{variation} in objects, which is the `opposite' of looking for invariants.  This cannot be taken too far, however: though ``twice opposite'' could mean ``the same,'' GIST and information complexity are certainly not the same.  The comparison of the uncertainty vectors and the structural manifolds above reveals that they are different.
  Part of the distinction is that while GIST searches for `invariants,' it makes no distinction among the non-invariants about their degree of variance, while our metric does.

 The information complexity metric has a \emph{min} versus \emph{mean} operator which naturally captures the paradigm-specific and general orders and settings.  Might this approach---replacing \emph{min} with \emph{mean}---be usefully applied to Boolean complexity and GIST to yield a prediction for the general order?  
Boolean complexity is rooted in finding the minimally complicated definition of a category, and counting its complexity: applying the logic of replacing \emph{min} with \emph{mean} would suggest finding all ways to express a category and taking the mean complexity of those expressions.  This is plausibly well-defined, in that the Boolean complexity starts with a maximally redundant list of the elements of a category and then reduces redundant elements: perhaps the complexity values at each step could be averaged.  However, as this relies on heuristics to find the minimum value, the mean value of the path to the the minimum would presumably be even more sensitive to the details of the heuristics chosen.
On the other hand, there is no natural analog to extend GIST to the general order/setting using this approach.
GIST relies on the search for invariants while ignoring dimensions, and the notion of minimization doesn't appear play any role, so there is no opportunity to even consider the replacement of \emph{min} with \emph{mean} to broaden its application.


\section{Results} 
\label{sec:results}

In this section, we consider the application of the metric to different classification learning tasks.  

In Subsection \ref{sub:shj_results}, we apply the metric to the SHJ tasks.  We consider both the paradigm-specific and general order contexts.  This involves comparison to human data collected by  \citet{nosofsky1994comparing} for the paradigm-specific context and \citet{nosofsky1996learning} for the general context. 
 In Subsection \ref{sub:beyond_shj_results}, we apply the metric to a larger range of tasks beyond SHJ. This involves comparison to human data collected by \citet{vigo2013gist}.  These data cover more classification learning tasks, but only in the paradigm-specific context.\footnote{Therefore in this section we predict the outcome of future experiments: namely, these classification learning tasks performed in the general context.}  We compare the performance of the information complexity metric to Feldman's Boolean complexity \citep{feldman2000} and Vigo's GIST \citep{vigo2013gist}.
 
 In both of these task applications, the comparison to human data takes two forms: a qualitative comparison, in which the ordering of difficulty predicted by the relevant version of our metric is compared to the difficulty ordering observed in the human data; and a quantitative comparison, in which values of the relevant metric are used to predict human classification error rates.  Both kinds of comparisons are important in this literature and date back to the original SHJ analysis \citep{shepard1961learning}.

 \begin{table}
 	\begin{center}
 		\begin{tabular}{|ccc||cccccc|}
 			\hline
 			\multicolumn{3}{|c||}{\textbf{Dim.}} & 
 			\multicolumn{6}{|c|}{\textbf{Category (A or B)}} \\ 
 			\multicolumn{3}{|c||}{\textbf{Values}} & 
 			\multicolumn{6}{|c|}{\textbf{By SHJ Type ($I-VI$)}} \\ \hline
 			\textbf{1} & \textbf{2} &  \textbf{3}	& \textbf{$I$}  &  \textbf{$II$}	& \textbf{$III$} &  \textbf{$IV$}	&	\textbf{$V$} &  \textbf{$VI$} \\ \hline
 			0 & 0 &  0  &  A  &  A	& A	  &  A	&	A &  A  \\
 			0 & 0 &  1  &  A  &  A	& A	  &  A	&	A &  B  \\
 			0 & 1 &  0  &  A  &  B	& A	  &  A	&	A &  B  \\
 			0 & 1 &  1  &  A  &  B	& B	  &  B	&	B &  A  \\
 			1 & 0 &  0  &  B  &  B	& B	  &  A	&	B &  B  \\
 			1 & 0 &  1  &  B  &  B	& A	  &  B	&	B &  A  \\
 			1 & 1 &  0  &  B  &  A	& B	  &  B	&	B &  A  \\
 			1 & 1 &  1  &  B  &  A	& B	  &  B	&	A &  B \\ \hline
 		\end{tabular}
 		\medskip

 		\begin{tabular}{rlc}
 		Paradigm-Specific Order: & $I < \, II \, \leq \enspace III \,,\, IV \,,\, V \, < VI$ & 	\hspace{23 mm} \\ 
 		General Order: & $I < IV < III < V < II < VI$ & 
 		\end{tabular}
 	\end{center}
 	\caption{The six mappings of three-digit binary strings to categories, Type I-VI.}
 	\label{tab:SHJmappings}
 \end{table}
 
\subsection{SHJ Tasks: Paradigm-specific versus General Orders} 
\label{sub:shj_results}

Consider the application to the SHJ tasks.  Table \ref{tab:SHJmappings} depicts the definitions of SHJ tasks $I$ through $VI$ as well as the paradigm-specific and general orders.  
In Subsection \ref{sub:our_metric}, we discussed two metrics, \uminhat{} and \umeanhat{}, and mentioned that \uminhat{} empirically predicts the paradigm-specific order and \umeanhat{} empirically predicts the general order.

\begin{table}
	\begin{center}
		\begin{tabular}{c|cccccc|l}																	
		&	\multicolumn{6}{c|}{	{SHJ Types}											} &	\multicolumn{1}{|c}{Order} \\	
		&\textbf{$I$}  &  \textbf{$II$}	& \textbf{$III$} &  \textbf{$IV$}	&	\textbf{$V$} &  \textbf{$VI$}	\\	\hline  & & & & & & &	\\
$\umin(0)$ &	1	&	1	&	1	&	1	&	1	&	1	&	\textit{	$I =\, II =\,III =^\star IV =^\star  V =\, VI$	} \\  & & & & & & & \\
$\umin(1)$ &	0	&	1	&	0.81	&	0.81	&	0.81	&	1	&	\textit{	$I <^\star III =^\star IV =^\star  V <\, II =\, VI$	} \\  & & & & & & & \\
$\umin(2)$ &	0	&	0	&	0.5	&	0.5	&	0.5	&	1	&	\textit{	$I =\, II <^\star III =^\star IV =^\star   V <^\star VI$	} \\  & & & & & & & \\
$\umin(3)$ &	0	&	0	&	0	&	0	&	0	&	0	&	\textit{	$I =\, II =\,III =^\star IV =^\star  V =\, VI$	} \\  & & & & & & & \\ \hline & & & & & & & \\
$\uminhat$ &	1	&	2	&	2.31	&	2.31	&	2.31 &	3	&	\textit{	$I <^\star  II<^\star III =^\star IV =^\star  V <^\star VI$	} 
		\end{tabular}
	\end{center}
	\caption{The information complexity metric \umin{} calculated for the SHJ classification learning problems.  The last line contains the aggregate metric and is the sum of the previous lines.  ``$\star$'' indicates that this relationship matches the paradigm-specific order.}
	\label{tab:paradigm-order-outcome}
\end{table}

\bigskip

\begin{table}
	\begin{center}
		\begin{tabular}{c|cccccc|l}																	
		&	\multicolumn{6}{c|}{	{SHJ Types}											} &	\multicolumn{1}{|c}{Order} \\	
		&\textbf{$I$}  &  \textbf{$II$}	& \textbf{$III$} &  \textbf{$IV$}	&	\textbf{$V$} &  \textbf{$VI$}	\\	\hline  & & & & & & &	\\
		$\umean(0)$		&	1	&	1	&	1	&	1	&	1	&	1	&	
					\textit{		    $I =\, IV =\, III =\, V =\, II =\, VI$							}	\\	 & & & & & & &	\\
		$\umean(1)$	&	0.67	&	1.00	&	0.87	&	0.81	&	0.94	&	1.00	&	
					\textit{		    $I <^\star IV <^\star III <^\star V <^\star II =\, VI$							}		\\ & & & & & & &	\\	
		$\umean(2)$		&	0.33	&	0.67	&	0.50	&	0.50	&	0.67	&	1.00	&	
					\textit{		        $I <^\star IV =\, III <^\star V =\, II <^\star VI$							}		\\ & & & & & & &	\\	
		$\umean(3)$		&	0	&	0	&	0	&	0	&	0	&	0 &
					\textit{		    $I =\, IV =\, III =\, V =\, II =\, VI$							}	\\  & & & & & & & \\ \hline & & & & & & & \\
		$\umeanhat$	&	2.00	&	2.67	&	2.37	&	2.31	&	2.60	&	3.00			
					&	\textit{		    $I <^\star IV <^\star III <^\star V <^\star II <^\star VI$							}	
		\end{tabular}
	\end{center}
	\caption{The information complexity metric \umean{} calculated on the SHJ classification learning problems.  The last line contains the aggregate metric and is the sum of the previous lines.   ``$\star$'' indicates that this relationship matches the general order.}
	\label{tab:general-order-outcome}
\end{table}

\emph{Qualitative Comparison.}
 Table \ref{tab:paradigm-order-outcome} gives the outcome of \umin{} metrics calculated for each SHJ task.  Because the SHJ stimuli have three dimensions, the first four rows give the intermediate metric values $\umin(0)$ through $\umin(3)$, followed by the aggregate information complexity metric $\uminhat$.  $\uminhat$ correctly predicts the paradigm-specific order.  To understand the application to SHJ, consider $\umin(1)$ and $\umin(2)$. 
  $\umin(1)$ finds $I		<	III, IV, V	<	II , VI$ and
   $\umin(2)$ finds $I , II	<	III, IV, V	<		VI$. 
  Neither is the full paradigm-specific order: they must be summed into $\uminhat$ in order to recover the paradigm-specific order. 
Why the sum?  Each rule extracts some amount of information, so the sum measures total information.
 Interestingly, all of the hallmarks of the SHJ paradigm-specific ordering are preserved in each case---except for Type $II$.  
  This potentially provides an account of individual differences that were observed in \citet{kurtzRevising}, which showed a bimodal distribution in Type II learning.\footnote{\citet{kurtzRevising} found that participants given task instructions to look for rules during learning were likely to show a Type II advantage---unlike those given neutral instructions.  We speculate that performance variation driven by task context or by individual differences may reflect processing of a limited or restricted nature.}  That is, some subjects learn Type $II$ very fast and some learn it quite slow, and aggregate results essentially reflect the weighting of the two types of learners in a sample.   It may be the case that subjects who are, implicitly or explicitly, constructing rules that focus on fixing one dimension are the ones who learn Type $II$ slowly, but those who are constructing rules that focus on fixing two dimensions are the ones who learn it quickly.
 It also might be that agents who focus on one rule over the other weight that information more heavily: this suggests that a metric like $\alpha \umin(1) + \left(1-\alpha\right)\umin(2)$ for  $\alpha \in \lbrack 0,1\rbrack$ which vary across people could account for individual performance.
  
The mean information complexity metric also displays this characteristic, that no single level fully captures the general order. Consider $\umean(1)$ and $\umean(2)$ from Table \ref{tab:general-order-outcome}. 
 $\umean(1)$ finds $II = VI$ and and $\umean(2)$ finds $IV = III$ and $V=II$.   Neither metric alone fully captures human learning difficulty in the general order: the general order is only recovered by summing to $\umeanhat{}.$

\begin{figure}[htbp]
	\centering
\includegraphics[scale=.36]{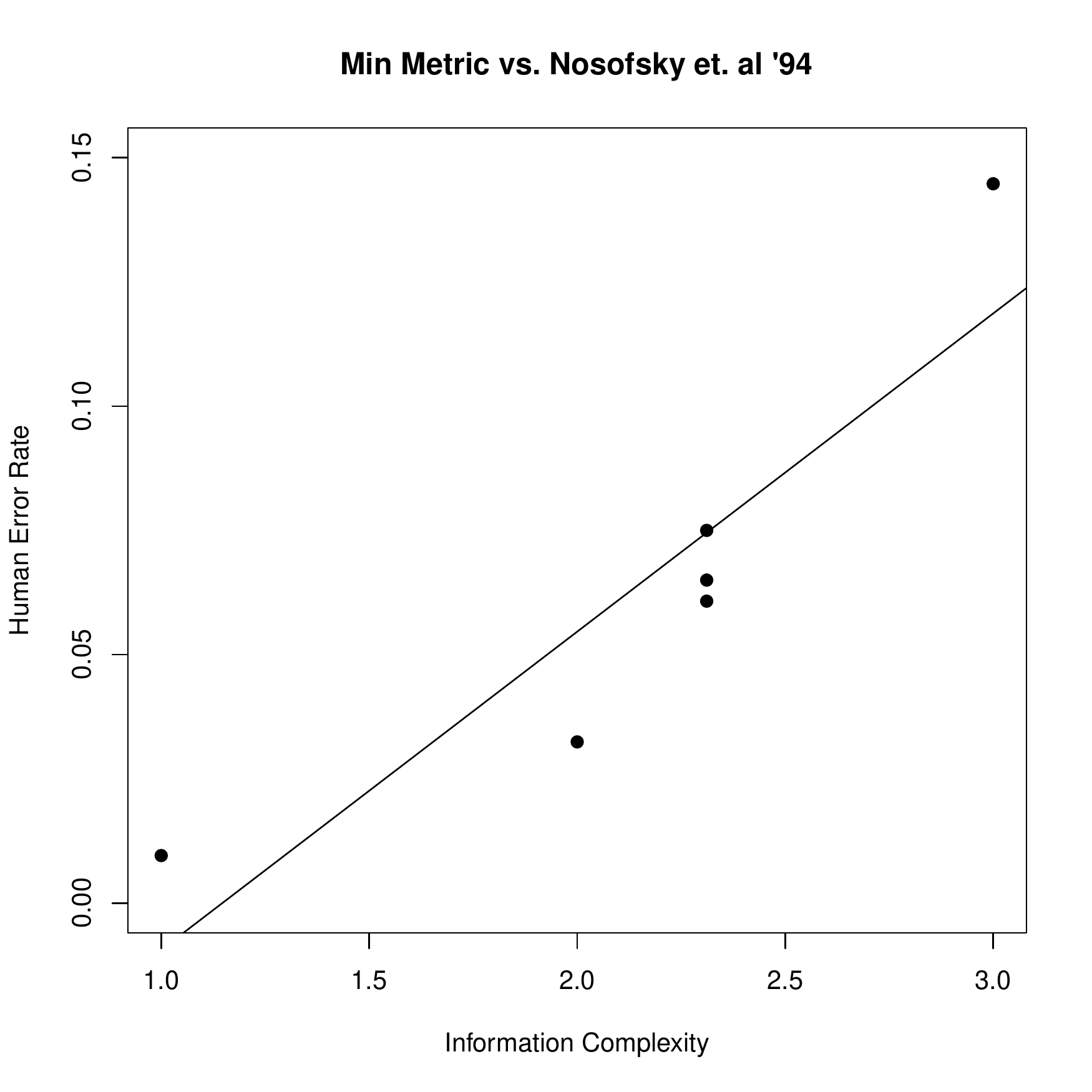}\qquad%
\includegraphics[scale=.36]{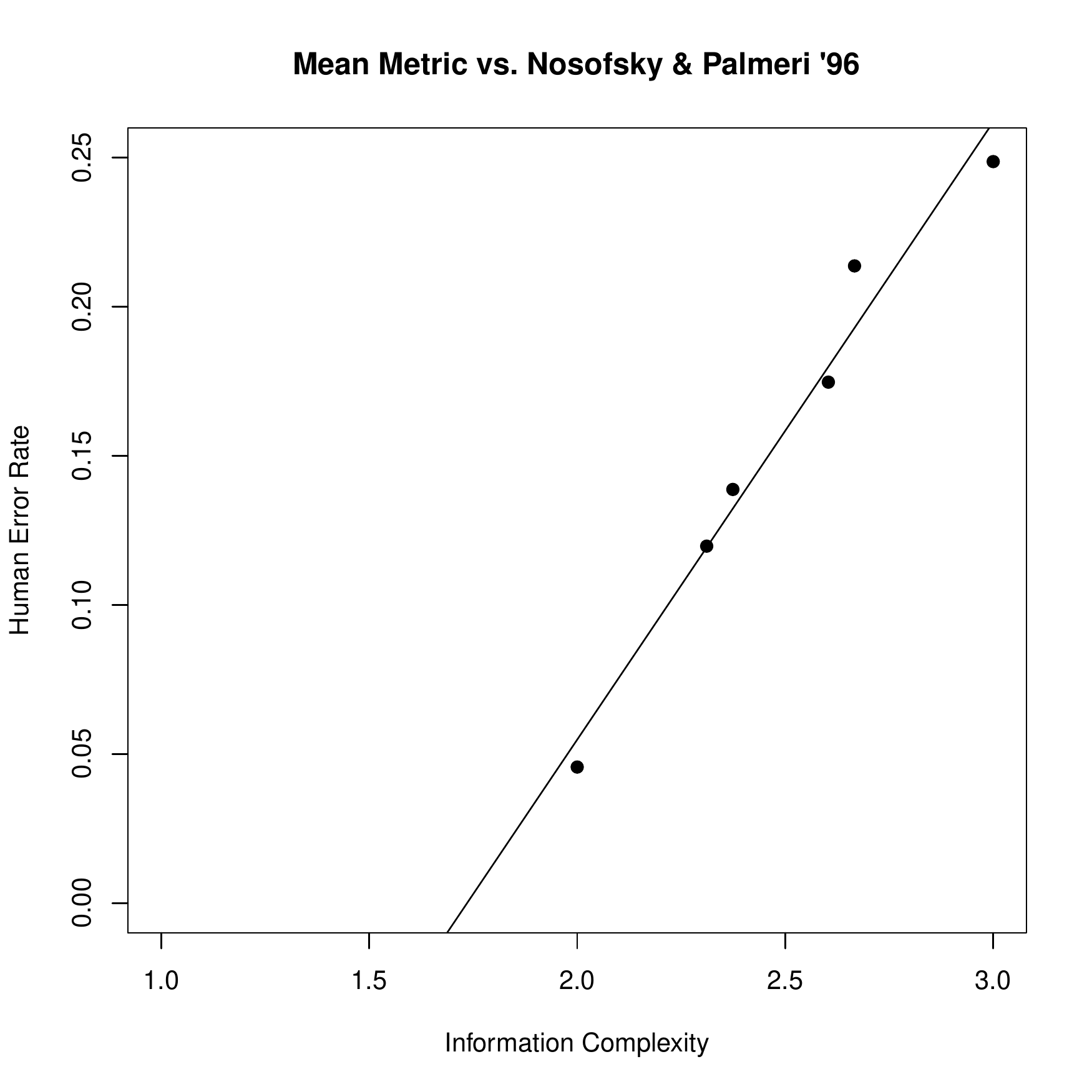}
	\caption{Quantitative Comparison: \uminhat{} and \umeanhat{} vs. Human Data (Nosofsky et. al '94 and Nosofsky \& Palmeri '96).}
	\label{fig:shj-quant-match}
\end{figure}

\bigskip

\emph{Quantitative Comparison.} Now, we consider the quantitative fit of the information complexity metric to two established data sets: data gathered by \citet{nosofsky1994comparing} in the paradigm-specific setting
and data gathered by 
\citet{nosofsky1996learning} in the general setting.  We average the probability of error across the 25 blocks of data for each SHJ Type, and then calculate the coefficient of determination ($R^2$) between the relevant information complexity metric and the average probability of error.  We find that the $R^2$ between the paradigm-specific human data and \uminhat{} is $.829$ and the $R^2$ between the general human data and \umeanhat{} is $0.970$.  The scatterplots with best-fit lines can be seen in Figure \ref{fig:shj-quant-match}.



\newsavebox{\savepar}
\newenvironment{boxit}{\begin{lrbox}{\savepar}
  }
  {\end{lrbox}\fbox{\usebox{\savepar}}}

\begin{table}
\begin{minipage}[c]{.6\textwidth}
\begin{boxit}
$2[1]$\\
$
\begin{array}{ccc}
 \text{A} & \hat{u}_{\min } & \hat{u}_{\text{mean}} \\
 \{00\} & 1.31 & 1.31 \\
\end{array}
$ \end{boxit}\\\\
\begin{boxit}
$2[2]$\\
$
\begin{array}{ccc}
 \text{A} & \hat{u}_{\min } & \hat{u}_{\text{mean}} \\
 \{00,01\} & 1. & 1.5 \\
 \{00,11\} & 2. & 2. \\
\end{array}
$ \end{boxit}\\\\
\bigskip
\\
\begin{boxit}
$3[1]$\\
$
\begin{array}{ccc}
 \text{A} & \hat{u}_{\min } & \hat{u}_{\text{mean}} \\
 \{000\} & 1.2 & 1.2 \\
\end{array}
$ \end{boxit}\\\\
\begin{boxit}
$3[2]$\\
$
\begin{array}{ccc}
 \text{A} & \hat{u}_{\min } & \hat{u}_{\text{mean}} \\
 \{000,001\} & 1.31 & 1.74 \\
 \{000,011\} & 1.81 & 2.02 \\
 \{000,111\} & 2.12 & 2.12 \\
\end{array}
$ \end{boxit}\\\\
\begin{boxit}
$3[3]$\\
$
\begin{array}{ccc}
 \text{A} & \hat{u}_{\min } & \hat{u}_{\text{mean}} \\
 \{000,001,010\} & 1.61 & 2.11 \\
 \{000,001,110\} & 2.11 & 2.44 \\
 \{000,011,101\} & 2.61 & 2.61 \\
\end{array}
$ \end{boxit}\\\\
\begin{boxit}
$3[4]$
\\
$
\begin{array}{r@{\,}lcc}
\multicolumn{2}{c}{\text{SHJ Type:\,A}} 			   		&	\hat{u}_{\min } & \hat{u}_{\text{mean}} \\
\text{I:}&\{000,001,010,011\}  &	1. & 2. \\
\text{II:}&\{000,001,110,111\}  &	2. & 2.67 \\
\text{III:}&\{000,001,010,101\}  &	2.31 & 2.37 \\
\text{IV:}&\{000,001,010,100\}  &	2.31 & 2.31 \\
\text{V:}&\{000,001,010,111\}  &	2.31 & 2.60 \\
\text{VI:}&\{000,011,101,110\}  &	3. & 3. \\
\end{array}
$ \end{boxit}
\end{minipage}%
\begin{minipage}[c]{.4\textwidth}
\begin{boxit}
$4[1]$\\
$
\begin{array}{ccc}
 \text{A} & \hat{u}_{\min } & \hat{u}_{\text{mean}} \\
 \{0000\} & 0.93 & 0.93 \\
\end{array}
$ \end{boxit}\\\\
\begin{boxit}
$4[2]$\\
$
\begin{array}{ccc}
 \text{A} & \hat{u}_{\min } & \hat{u}_{\text{mean}} \\
 \{0000,0001\} & 1.2 & 1.5 \\
 \{0000,0011\} & 1.45 & 1.64 \\
 \{0000,0111\} & 1.61 & 1.71 \\
 \{0000,1111\} & 1.74 & 1.74 \\
\end{array}
$ \end{boxit}\\\\
\begin{boxit}
$4[3]$\\
$
\begin{array}{ccc}
 \text{A} & \hat{u}_{\min } & \hat{u}_{\text{mean}} \\
 \{0000,0001,0010\} & 1.5 & 1.97 \\
 \{0000,0001,0110\} & 1.75 & 2.14 \\
 \{0000,0001,1110\} & 1.95 & 2.22 \\
 \{0000,0011,0101\} & 2.01 & 2.24 \\
 \{0000,0011,1100\} & 2.21 & 2.32 \\
 \{0000,0011,1101\} & 2.21 & 2.34 \\
\end{array}
$ \end{boxit}\\\\
\begin{boxit}
$4[4]$\\
$
\begin{array}{ccc}
 \text{A} & \hat{u}_{\min } & \hat{u}_{\text{mean}} \\
 \{0000,0001,0010,0011\} & 1.31 & 2.19 \\
 \{0000,0001,0010,0100\} & 1.97 & 2.34 \\
 \{0000,0001,0010,0101\} & 1.97 & 2.37 \\
 \{0000,0001,0010,0111\} & 1.97 & 2.5 \\
 \{0000,0001,0010,1100\} & 2.22 & 2.55 \\
 \{0000,0001,0010,1101\} & 2.22 & 2.56 \\
 \{0000,0001,0010,1111\} & 2.22 & 2.61 \\
 \{0000,0001,0110,0111\} & 1.81 & 2.52 \\
 \{0000,0001,0110,1010\} & 2.47 & 2.67 \\
 \{0000,0001,0110,1011\} & 2.47 & 2.71 \\
 \{0000,0001,0110,1110\} & 2.31 & 2.6 \\
 \{0000,0001,0110,1111\} & 2.31 & 2.73 \\
 \{0000,0001,1110,1111\} & 2.12 & 2.66 \\
 \{0000,0011,0101,0110\} & 2.31 & 2.7 \\
 \{0000,0011,0101,1001\} & 2.72 & 2.72 \\
 \{0000,0011,0101,1010\} & 2.72 & 2.77 \\
 \{0000,0011,0101,1110\} & 2.72 & 2.84 \\
 \{0000,0011,1100,1111\} & 2.62 & 2.83 \\
 \{0000,0011,1101,1110\} & 2.62 & 2.88 \\
\end{array}
$ \end{boxit}
\end{minipage}
		\caption{Results of the Information Complexity Metric $\hat{u}$}
	\label{tab:sweep}
\end{table}

\subsection{Beyond SHJ} 
\label{sub:beyond_shj_results}

Now we consider the application of the information complexity metric beyond the SHJ category structures.  
In \citet{feldman2000,feldman2003catalog}, he identified the extended catalog of logical category structures beyond SHJ.  In the notation he introduced,  a set of tasks $D[P]$ indicates objects of $D$ dimensions with $P$ objects in category $A$.  In this notation, the tasks we consider here are: $2[1]$, $2[2]$, $3[1]$, $3[2]$, $3[3]$, $3[4]$ (SHJ), $4[1]$, $4[2]$, $4[3]$ and $4[4]$. 
 We find the quantitative $R^2$ fit and qualitative ordering fit  to human data collected by \citet{vigo2013gist}.  We compare this fit against other metrics: Vigo's GIST \citep{vigo2013gist} and Feldman's Boolean complexity \citep{feldman2000}. 
\citet{vigo2013gist} represents an important advance in the literature on complexity-based accounts of category learning: 
(1) the GIST account is an alternative to Boolean algebraic rules based on the core principle of deriving complexity from the degree of categorical invariance to perturbing transformations;   
(2) a wide set of human learning data on logical category structures are provided using an improved methodology over what was previously available in this domain;\footnote{%
Vigo's description of his improved methodology over \citet{feldman2000}, 
	which we find compelling, describes four improvements: 
	the same time for subjects to learn stimuli across tasks, regardless of dimensionality; %
	sampling from all possible structures consistent with a particular $D[P]$ value, instead of a subset; %
	stimuli that are less abstract (images of flasks instead of `amoeba'); %
	and
	assigning stimuli to subjects in a way that all experiments took roughly the same time, 
	which would reduce errors due to subject fatigue.}  
%
and (3) impressive fits are demonstrated between the two.

Table \ref{tab:sweep} shows the values of \uminhat{} and \umeanhat{} for the tasks $2[1]$, $2[2]$, $3[1]$, $3[2]$, $3[3]$, $3[4]$ (SHJ), $4[1]$, $4[2]$, $4[3]$ and $4[4]$.  The first column, $A$, explicitly lists the elements of a representative $A$ set, when the category in question is viewed in ``up parity;'' i.e. when category $A$ has fewer elements than its complementary category $B$.  For example, $3[4]$, the SHJ tasks, lists the first set $A$ as $\left\{000,001,010,011\right\}$, which is the familiar Type $I$ problem, in which the first dimension is sufficient to determine whether a particular stimulus is in category $A$.  (Note that, due to symmetry, there is no distinction between up and down parity in the SHJ tasks.)   For the convenience of the reader, in Table  \ref{tab:sweep} the tasks in each category structure are listed in the same order that they appear elsewhere \citep{feldman2000,feldman2003catalog,vigo2013gist}.

\bigskip
	\begin{table}[htbp]
		\begin{center}  
		\begin{tabular}{l||c|c|c}
		&	\uminhat{}	&	GISTM	& BoolC\\ \hline & & & \\
	3[2]										&	{1}		& {.866}	  &  1  \\ & &\\
	3[3]										&	{1}		& {1}		  &  1  \\ & &\\
	3[4]  \emph{\footnotesize (SHJ)}	&	{.941}	& {.812}	  &  .941  \\ & &\\
	4[2]										& 	{.8} 		& {1}		  &  .8  \\ & &\\
	4[3]										&	{.986}	& {.926}	  &  .986  \\ & &\\
	4[4]										&	{.910}	& {.860}	  &  .789  \\ & &\\ \hline
		\end{tabular}
		\end{center}
		\caption{Spearman $\rho$: Information Complexity vs. GIST vs. Boolean Complexity. The value indicates correlation in orders between the metric and human data.}
		\label{tab:rank-order}
	\end{table}

\emph{Qualitative Comparison.}
First, we consider the relative difficulty orderings implied by Feldman's Boolean complexity \citep{feldman2003catalog}, Vigo's GISTM \citep{vigo2013gist}, and \uminhat{}, and compare them to the orderings found in Vigo's data.%
\footnote{
The orderings implied by Vigo's GISTM metric are extrapolated from the values of $\hat{\Phi}$ given in \citet{vigo2013gist} Table 1.  
 Since GISTM is the `core model' of Vigo's approach, we believe that the incorporation of GISTM orderings are sufficient for this analysis.} 
The data in question are from the 84 structures tested Vigo's Experiment 1, what Vigo refers to as VEXPRO-84 in his text.\footnote{We would like to thank Dr. Vigo for making these data available to us for this analysis.}  
In this experiment, human adult learners (Ohio University undergraduates) are presented with stimuli with separable dimensions (images of flasks which varied in color, size, shape, and neck width). 
Since this experiment is in the paradigm-specific setting, \uminhat{} is our preferred metric.

The rank correlations, also called Spearman $\rho$, between the data and the available metrics are depicted in Table \ref{tab:rank-order}.  They provide a quantitative metric to measure ordering accuracy.  As can be seen, \uminhat{} provides the equal or better rank correlation for each category structures except for $4[2]$, where GISTM provides the best rank correlation.  The different category structures are discussed in detail below.
 
There is only one non-trivial case for two dimensions, which is $2{[}2{]}$.  For these category structures, Boolean complexity, GISTM, and  \uminhat{} orderings match, but no human data are available.

There are three non-trivial cases for three dimensions, and differences across metrics and the human data are observed.
For $3{[2]}$, Boolean complexity and \uminhat{} orderings match each other and also match the order observed in the data: $\{000,001\} < \{000,011\} < \{000,111\}$.  However, the GISTM predicts that  $\{000,001\}$  and $\{000,111\}$ are of equal difficulty, which is not observed in the data.  As can be seen in Table \ref{tab:rank-order}, this causes GISTM to have a lower rank correlation with the human data.
For $3{[3]}$, Boolean complexity, GISTM, \uminhat{}, and the observed data orderings all match, and therefore perfect rank correlations are observed in Table \ref{tab:rank-order}.

$3{[4]}$ are the SHJ tasks, discussed above, where the Boolean complexity ordering matches the \uminhat{} ordering. GISTM ordering is similar to Boolean complexity and \uminhat{} ordering, except that $III = IV < V$ instead of $III=IV=V$.  Note that the correlations in Table \ref{tab:rank-order} are not $1$ because there are actually slight differences in the human data between $III$, $IV$, and $V$, and rank order is strict with regards to those slight differences.  And GISTM's prediction over those three items does run contrary to the small human variation.
(Also, as noted above, neither GISTM nor Boolean complexity predict the general order, which we predict with \umeanhat{}.  See Section \ref{sub:shj_results}).

There are also three non-trivial cases for four dimensions that we consider here.
For $4{[2]}$, Boolean complexity and \uminhat{} match each other:
\begin{equation} 
 \{0000,0001\} < \{0000,0011\} < \{0000,0111\} < \{0000,1111\} \nonumber
\end{equation} 
In the data, we observe a reversal of the last two categories, namely that $\{0000,1111\} < \{0000,0111\}$.
GISTM predicts that $\{0000,0111\}$ and $\{0000,1111\}$ are of equal difficulty.  This causes GISTM to have a higher rank correlation for $4[2]$, as observed in Table \ref{tab:rank-order}.

For $4{[3]}$, Boolean complexity and \uminhat{} orderings match each other: there is a strict ordering over the first five tasks, and the fifth and sixth tasks are equally difficult.  This very nearly matches the human data: in the human data three is a strict order over all items, including the last two (the sixth is more difficult than the fifth.)  GISTM predicts many ties between items: GISTM predicts the second and third items are of equal difficulty, and predicts that the fourth, fifth, and sixth items are of equal difficulty.  This leads GISTM to have a lower rank correlation than does \uminhat{}.

Finally, we consider the most complicated set of category structures, $4{[4]}$.
All orderings differ from each other, and all differ from the data.  Boolean complexity and \uminhat{} match fairly well, although \uminhat{} finds two pairs to be identical that Boolean complexity differentiates.  \uminhat{} also finds that the $11^\textrm{th}$ item, 
$\{0000, 0001, 0110, 1110\}$, is moderately difficult, consistent with the human data, while Boolean complexity finds it fairly easy.  The combination of these factors cause Boolean complexity to have a relatively low $\rho$ value.
  GISTM declares many cases to be identical which other metrics differentiate (two pairs of identical tasks, two triples of identical tasks, and one group of six identical tasks.)  This also lowers GISTM's match to human rankings relative to \uminhat{}.

\bigskip
\begin{table}[htbp]
	\begin{center}
	\begin{tabular}{l||cc|cc|c}
&	\uminhat{}	&	\umeanhat{}	&	GISTM-SE	&	GISTM	&	BoolC	\\ 
	\hline & & & & & \\
3[2]	& \textbf{	0.994	} &	0.999	& {	0.950	} &	0.910	& {	0.995	} \\ & & & & & \\
3[3]	& \textbf{	0.930	} &	0.805	& {	0.890	} &	0.925	& {	0.930	} \\ & & & & & \\
3[4]	& \textbf{	0.877	} &	0.519	& {	1.000	} &	0.920	& {	0.965	} \\ & & & & & \\
4[2]	& \textbf{	0.765	} &	0.873	& {	0.915	} &	0.835	& {	0.912	} \\ & & & & & \\
4[3]	& \textbf{	0.891	} &	0.866	& {	0.945	} &	0.945	& {	0.978	} \\ & & & & & \\
4[4]	& \textbf{	0.915	} &	0.913	& {	0.850	} &	0.845	& {	0.686	} \\  & & & & & \\\hline
	\end{tabular}
	\end{center}
	\caption{Coefficient of Determination $R^2$: Information Complexity vs. GIST vs. BoolC.  Because this is a paradigm-specific setting, \uminhat{} is the preferred information complexity metric (indicated in bold).}
	\label{tab:datacomparison}
\end{table}

\emph{Quantitative Comparison.}
Table \ref{tab:datacomparison} depicts the coefficient of determination ($R^2$) of the information complexity metrics, Vigo's metrics of GIST-SE and GIST-M, and Feldman's Boolean complexity 
against Vigo's human data. 
The GISTM-SE and GISTM $R^2$ values depicted here are from Vigo's Figure 7, averaged over up and down parity \citep{vigo2013gist}.\footnote{We do this because our metric does not differentiate between up and down parity, because it is symmetric with regards to included and excluded categories.  For the same reason, we collapse the behavioral data collected by Vigo over parity as well.  It is worth noting that the up-down parity differences are external to the core ordering effects.}  
The Boolean complexity $R^2$ values are calculated by the authors using the values for Boolean complexity given in \citet{feldman2003catalog}.

\uminhat{} should provide a better fit to these data than \umeanhat{}.  We find this to be the case in $3[3]$, $3[4]$ (previously noted), $4[3]$, and $4[4]$, which is consistent with this hypothesis.  Before discussing the two exceptions, $3[2]$ and $4[2]$, below, we wish to note that this table also yields a prediction for future experiments: we predict that experiments $3[3]$, $4[3]$, and $4[4]$ performed in the general setting---i.e. with integral dimensions, with children or monkeys as subjects, or  when categorization difficulty is extrapolated from errors in identification learning---should be better explained by \umeanhat{} than \uminhat{}.  (Please see Table \ref{tab:sweep} for those values). 

For $3[2]$ and $4[2]$, we find \umeanhat{} provides a better fit than \uminhat{}. The explanation we find most likely is illuminating.  
The explanation that \uminhat{} ought to be a better fit than \umeanhat{} requires that subjects either implicitly or explicitly search for dimensions that explain the categorization best.  
In $3[2]$, category $A$ is only two objects from $2^3 = 8$ total; and in $4[2]$, category $A$ is only two objects from $2^4= 16$ total. 
No matter how those two stimuli are distributed in the stimulus space, there is hardly any meaningful structure of subdimensions. 
This might be why our \uminhat{} metric does not perform well in this case. 
That is, subjects are not searching for dimension-based definitions of $A$, instead they may be memorizing the elements of $A$.
If this explanation is correct, it would also arise in any other $D[P]$ case where $D$ is large while $P$ is small. 
Moreover, if this explanation is correct, we would expect this effect to be more pronounced for $4[2]$ over $3[2]$ (because the set $A$ is smaller relative to the total number of objects); and indeed, $4[2]$ shows a more distinct advantage of \umeanhat{} over \uminhat{} than does $3[2]$, where the \umeanhat{} advantage is slight.

Consider the fit of the information complexity metrics relative to GISTM and GISTM-SE, and then to Boolean complexity, also depicted in Table \ref{tab:datacomparison}. 
\uminhat{} is a better fit than the better of GISTM-SE and GISTM in the cases of 
	$3[2]$, $3[3]$, and $4[4]$; 
	the reverse holds for $3[4]$, $4[2]$, and $4[3]$.  
	From this perspective, the metrics are of comparable quantitative fit.  However, it should be noted that the GIST metrics involve a free parameter ($k$) fitted to data separately for each set of category structures, which a degree of freedom the information complexity metric does not have.   \uminhat{} is a comparable fit to Boolean complexity in $3[2]$ and $3[3]$,\footnote{Although \umeanhat{} is a better fit than Boolean complexity for $3[2]$.  See discussion above on that point.} a worse fit in $3[4]$, $4[2]$, and $4[3]$, and a substantially better fit in $4[4]$.  So the quantitative fit comparison here is mixed.


\section{Conclusions} 
\label{sec:conclusion}

The existing complexity metric literature concludes that ``human conceptual difficulty reflects intrinsic mathematical complexity[.]''\citep{feldman2000}  Our finding strengthens and deepens this fundamental result.  We find that human conceptual difficulty reflects information complexity, where information complexity, based on Shannon entropy, is a concept based on uncertainty remaining as dimensions or sets of dimensions are specified.  This approach has the advantage of being able to predict human learning whether in the paradigm-specific setting---in which adult learners are shown objects with separable dimensions---or in the general setting---in which dimensions are integral, in which learners are children or monkeys, or if errors are extrapolated from identification learning.  Importantly, the second setting has never before been predicted by any mathematical complexity approach.  Moreover, this approach explains these two domains in a way that reveals a possibly deep connection: when dimensions are separable, human learners are able to identify the dimension or set of dimensions that yield the minimum uncertainty, and exploit that; and when learners do not or cannot differentiate dimensions, then average uncertainty best predicts human learning.  Finally, our metric makes predictions about the difficulty ordering for classification learning experiments beyond SHJ in the general setting. As experiments about general-setting learning beyond SHJ accumulate, we will learn whether and how effectively this measure of mathematical complexity can predict learning behavior.


\clearpage
\newpage 

\bibliographystyle{model5-names}
\bibliography{complexity-concept}

\begin{thebibliography}{23}
\expandafter\ifx\csname natexlab\endcsname\relax\def\natexlab#1{#1}\fi
\providecommand{\bibinfo}[2]{#2}
\ifx\xfnm\relax \def\xfnm[#1]{\unskip,\space#1}\fi
\bibitem[{Feldman(2000)}]{feldman2000}
\bibinfo{author}{Feldman, J.} (\bibinfo{year}{2000}).
\newblock \bibinfo{title}{{Minimization of Boolean complexity in human concept
  learning}}.
\newblock {\it \bibinfo{journal}{Nature}\/},  {\it \bibinfo{volume}{407}\/},
  \bibinfo{pages}{630--633}.
\bibitem[{Feldman(2003)}]{feldman2003catalog}
\bibinfo{author}{Feldman, J.} (\bibinfo{year}{2003}).
\newblock \bibinfo{title}{A catalog of boolean concepts}.
\newblock {\it \bibinfo{journal}{Journal of Mathematical Psychology}\/},  {\it
  \bibinfo{volume}{47}\/}, \bibinfo{pages}{75--89}.
\bibitem[{Feldman(2006)}]{feldman2006algebra}
\bibinfo{author}{Feldman, J.} (\bibinfo{year}{2006}).
\newblock \bibinfo{title}{An algebra of human concept learning}.
\newblock {\it \bibinfo{journal}{{Journal of Mathematical Psychology}}\/},
  {\it \bibinfo{volume}{50}\/}, \bibinfo{pages}{339--368}.
\bibitem[{Garner(1974)}]{garner1974processing}
\bibinfo{author}{Garner, W.~R.} (\bibinfo{year}{1974}).
\newblock {\it \bibinfo{title}{{The Processing of Information and
  Structure}}\/}.
\newblock \bibinfo{address}{{Potomac, MD}}: \bibinfo{publisher}{Lawrence
  Erlbaum}.
\bibitem[{Goodman et~al.(2008)Goodman, Tenenbaum, Feldman \&
  Griffiths}]{Goodman2008}
\bibinfo{author}{Goodman, N.~D.}, \bibinfo{author}{Tenenbaum, J.~B.},
  \bibinfo{author}{Feldman, J.}, \& \bibinfo{author}{Griffiths, T.~L.}
  (\bibinfo{year}{2008}).
\newblock \bibinfo{title}{A rational analysis of rule-based concept learning}.
\newblock {\it \bibinfo{journal}{Cognitive Science}\/},  {\it
  \bibinfo{volume}{32}\/}.
\bibitem[{Goodwin \& Johnson-Laird(2011)}]{goodwin2011mental}
\bibinfo{author}{Goodwin, G.~P.}, \& \bibinfo{author}{Johnson-Laird, P.}
  (\bibinfo{year}{2011}).
\newblock \bibinfo{title}{{Mental models of Boolean concepts}}.
\newblock {\it \bibinfo{journal}{{Cognitive Psychology}}\/},  {\it
  \bibinfo{volume}{63}\/}, \bibinfo{pages}{34--59}.
\bibitem[{Kruschke(1992)}]{kruschke1992alcove}
\bibinfo{author}{Kruschke, J.} (\bibinfo{year}{1992}).
\newblock \bibinfo{title}{{ALCOVE: An exemplar-based connectionist model of
  category learning}}.
\newblock {\it \bibinfo{journal}{{Psychological Review}}\/},  {\it
  \bibinfo{volume}{99}\/}, \bibinfo{pages}{22}.
\bibitem[{Kurtz(2007)}]{kurtz2007divergent}
\bibinfo{author}{Kurtz, K.~J.} (\bibinfo{year}{2007}).
\newblock \bibinfo{title}{{The divergent autoencoder (DIVA) model of category
  learning}}.
\newblock {\it \bibinfo{journal}{{Psychonomic Bulletin \& Review}}\/},  {\it
  \bibinfo{volume}{14}\/}, \bibinfo{pages}{560--576}.
\bibitem[{Kurtz et~al.(2013)Kurtz, Levering, Romero, Stanton \&
  Morris}]{kurtzRevising}
\bibinfo{author}{Kurtz, K.~J.}, \bibinfo{author}{Levering, K.},
  \bibinfo{author}{Romero, J.}, \bibinfo{author}{Stanton, R.~D.}, \&
  \bibinfo{author}{Morris, S.~N.} (\bibinfo{year}{2013}).
\newblock \bibinfo{title}{{Human learning of elemental category structures:
  Revising the classic result of Shepard, Hovland, and Jenkins (1961)}}.
\newblock {\it \bibinfo{journal}{{Journal of Experimental Psychology: Learning,
  Memory, and Cognition}}\/},  {\it \bibinfo{volume}{39}\/},
  \bibinfo{pages}{552--572}.
\bibitem[{Lafond et~al.(2007)Lafond, Lacouture \&
  Mineau}]{lafond2007complexity}
\bibinfo{author}{Lafond, D.}, \bibinfo{author}{Lacouture, Y.}, \&
  \bibinfo{author}{Mineau, G.} (\bibinfo{year}{2007}).
\newblock \bibinfo{title}{{Complexity minimization in rule-based category
  learning: Revising the catalog of Boolean concepts and evidence for
  non-minimal rules}}.
\newblock {\it \bibinfo{journal}{Journal of Mathematical Psychology}\/},  {\it
  \bibinfo{volume}{51}\/}, \bibinfo{pages}{57--74}.
\bibitem[{Li \& Vit{\^a}anyi(2008)}]{li2008introduction}
\bibinfo{author}{Li, M.}, \& \bibinfo{author}{Vit{\^a}anyi, P.}
  (\bibinfo{year}{2008}).
\newblock {\it \bibinfo{title}{An introduction to Kolmogorov complexity and its
  applications, Third Edition}\/}.
\newblock \bibinfo{publisher}{Springer}.
\bibitem[{Love et~al.(2004)Love, Medin \& Gureckis}]{love2004sustain}
\bibinfo{author}{Love, B.}, \bibinfo{author}{Medin, D.}, \&
  \bibinfo{author}{Gureckis, T.} (\bibinfo{year}{2004}).
\newblock \bibinfo{title}{{SUSTAIN: A network model of category learning}}.
\newblock {\it \bibinfo{journal}{{Psychological Review}}\/},  {\it
  \bibinfo{volume}{111}\/}, \bibinfo{pages}{309--332}.
\bibitem[{Minda et~al.(2008)Minda, Desroches \& Church}]{minda2008learning}
\bibinfo{author}{Minda, J.~P.}, \bibinfo{author}{Desroches, A.~S.}, \&
  \bibinfo{author}{Church, B.~A.} (\bibinfo{year}{2008}).
\newblock \bibinfo{title}{Learning rule-described and non-rule-described
  categories: A comparison of children and adults}.
\newblock {\it \bibinfo{journal}{{Journal of Experimental Psychology: Learning,
  Memory, and Cognition}}\/},  {\it \bibinfo{volume}{34}\/},
  \bibinfo{pages}{1518}.
\bibitem[{Nosofsky(1986)}]{nosofsky1986attention}
\bibinfo{author}{Nosofsky, R.} (\bibinfo{year}{1986}).
\newblock \bibinfo{title}{Attention, similarity, and the
  identification--categorization relationship}.
\newblock {\it \bibinfo{journal}{{Journal of Experimental Psychology:
  General}}\/},  {\it \bibinfo{volume}{115}\/}, \bibinfo{pages}{39--57}.
\bibitem[{Nosofsky et~al.(1994)Nosofsky, Gluck, Palmeri, McKinley \&
  Glauthier}]{nosofsky1994comparing}
\bibinfo{author}{Nosofsky, R.}, \bibinfo{author}{Gluck, M.},
  \bibinfo{author}{Palmeri, T.}, \bibinfo{author}{McKinley, S.}, \&
  \bibinfo{author}{Glauthier, P.} (\bibinfo{year}{1994}).
\newblock \bibinfo{title}{{Comparing models of rule-based classification
  learning: A replication and extension of Shepard, Hovland, and Jenkins
  (1961)}}.
\newblock {\it \bibinfo{journal}{{Memory and Cognition}}\/},  {\it
  \bibinfo{volume}{22}\/}, \bibinfo{pages}{352--352}.
\bibitem[{Nosofsky \& Palmeri(1996)}]{nosofsky1996learning}
\bibinfo{author}{Nosofsky, R.}, \& \bibinfo{author}{Palmeri, T.}
  (\bibinfo{year}{1996}).
\newblock \bibinfo{title}{{Learning to classify integral-dimension stimuli}}.
\newblock {\it \bibinfo{journal}{{Psychonomic Bulletin and Review}}\/},  {\it
  \bibinfo{volume}{3}\/}, \bibinfo{pages}{222--226}.
\bibitem[{Pape \& Kurtz(2013)}]{PapeKurtzGEB}
\bibinfo{author}{Pape, A.~D.}, \& \bibinfo{author}{Kurtz, K.~J.}
  (\bibinfo{year}{2013}).
\newblock \bibinfo{title}{Evaluating case-based decision theory: Predicting
  empirical patterns of human classification learning}.
\newblock {\it \bibinfo{journal}{Games and Economic Behavior}\/},  {\it
  \bibinfo{volume}{82}\/}, \bibinfo{pages}{52--65}.
\bibitem[{Shannon(1948)}]{shannon2001mathematical}
\bibinfo{author}{Shannon, C.~E.} (\bibinfo{year}{1948}).
\newblock \bibinfo{title}{A mathematical theory of communication}.
\newblock {\it \bibinfo{journal}{ACM SIGMOBILE Mobile Computing and
  Communications Review}\/},  {\it \bibinfo{volume}{5}\/},
  \bibinfo{pages}{3--55}.
\newblock \bibinfo{note}{Reprint published in 2001.}
\bibitem[{Shepard et~al.(1961)Shepard, Hovland \&
  Jenkins}]{shepard1961learning}
\bibinfo{author}{Shepard, R.}, \bibinfo{author}{Hovland, C.}, \&
  \bibinfo{author}{Jenkins, H.} (\bibinfo{year}{1961}).
\newblock \bibinfo{title}{Learning and memorization of classifications}.
\newblock {\it \bibinfo{journal}{{Psychological Monographs}}\/},  {\it
  \bibinfo{volume}{75}\/}, \bibinfo{pages}{1--41}.
\bibitem[{Smith et~al.(2004)Smith, Minda \& Washburn}]{smith2004category}
\bibinfo{author}{Smith, J.~D.}, \bibinfo{author}{Minda, J.~P.}, \&
  \bibinfo{author}{Washburn, D.~A.} (\bibinfo{year}{2004}).
\newblock \bibinfo{title}{{Category learning in rhesus monkeys: A study of the
  Shepard, Hovland, and Jenkins (1961) tasks.}}
\newblock {\it \bibinfo{journal}{{Journal of Experimental Psychology:
  General}}\/},  {\it \bibinfo{volume}{133}\/}, \bibinfo{pages}{398--414}.
\bibitem[{Vigo(2006)}]{vigo2006note}
\bibinfo{author}{Vigo, R.} (\bibinfo{year}{2006}).
\newblock \bibinfo{title}{{A note on the complexity of Boolean concepts}}.
\newblock {\it \bibinfo{journal}{{Journal of Mathematical Psychology}}\/},
  {\it \bibinfo{volume}{50}\/}, \bibinfo{pages}{501--510}.
\bibitem[{Vigo(2009)}]{Vigo2009}
\bibinfo{author}{Vigo, R.} (\bibinfo{year}{2009}).
\newblock \bibinfo{title}{Categorical invariance and structural complexity in
  human concept learning}.
\newblock {\it \bibinfo{journal}{Journal of Mathematical Psychology}\/},  {\it
  \bibinfo{volume}{53}\/}, \bibinfo{pages}{203--221}.
\bibitem[{Vigo(2013)}]{vigo2013gist}
\bibinfo{author}{Vigo, R.} (\bibinfo{year}{2013}).
\newblock \bibinfo{title}{{The GIST of concepts}}.
\newblock {\it \bibinfo{journal}{Cognition}\/},  {\it \bibinfo{volume}{129}\/},
  \bibinfo{pages}{138--162}.

\end{thebibliography}

\end{document}